\documentclass[10pt,a4paper]{iopart}
\usepackage{iopams}  
\usepackage{graphicx,psfrag,bbm,latexsym,color,dcolumn,bm,dsfont,bbm,color,
mathrsfs,bbold,latexsym,amsfonts,amssymb}

\def\Arctanh{\mathop{\rm \tanh^{-1}}\nolimits}

\def\t{\theta}
\def\s{\sigma}

\begin{document}

\title[Phase diagram of an Ising model with competitive interactions]
{Phase diagram of an Ising model with competitive
interactions on a Husimi tree and its disordered counterpart}
\author{Massimo Ostilli$^{1,2}$, F. Mukhamedov$^{1}$, J. F. F. Mendes$^{1}$}
\address{$^1$ Departamento de 
F{\'\i}sica da Universidade de Aveiro, 3810-193 Aveiro,
Portugal}
\address{$^2$ Center for Statistical Mechanics and Complexity (INFM-CNR), Italy}

\begin{abstract}
We consider an Ising competitive model defined over a triangular Husimi tree
where loops, responsible for an explicit frustration, are even allowed.
After a critical analysis of the phase diagram, in which a ``gas of non
interacting dimers (or spin liquid) - ferro or antiferromagnetic ordered state''
transition is recognized in the frustrated regions,
we introduce the disorder for studying the spin glass
version of the model: the triangular $\pm J$ model.
We find out that, 
for any finite value of the averaged couplings,
the model exhibits always a phase transition, 
even in the frustrated regions,
where the transition turns out to be a glassy transition.
The analysis of the random model 
is done by applying a recently proposed method
which allows to derive the upper phase boundary of a random
model through a mapping with a corresponding non random one. 
\end{abstract}

\ead{ostilli@roma1.infn.it}
\maketitle

\section{Introduction} \label{intro}

Models of statistical mechanics defined over Bethe lattices \cite{Baxter}
constitute a framework that, due to two peculiar ingredients,
namely exact solvability and finite connectivity, as opposed to
unsolvability of finite dimensional models (for $D>2$) 
or to exact solvability
of models defined over the fully connected graph (mean field),
sheds some light toward the understanding of more realistic
models. It is in fact believed that several thermal properties of
the model could persist for regular lattices as well. 
Nowadays, the emerging science of networks has attracted a
renewed interested in such models \cite{DM,Vespignani,Barbasi,DM1}.
In fact, roughly
speaking one can say that the Bethe lattice represents the
simplest prototype of network where an exact solution is often
accessible. At least for uncorrelated random networks,
it has been indeed proved that, if we indicate with $<f(q)>$ the
average over the network of some function $f$ of the vertex degree $q$,
the critical behavior of a given model with positive
couplings defined over a Bethe lattice of degree $q_B$,
coincides with that defined over the network provided that
an effective substitution
$q_B\rightarrow <q^2>/<q>$ for taking into account of the
distribution of link ends rather than that of links,
is performed \cite{Goltsev,Zecchina}.

\begin{figure}
\includegraphics[width=0.6\columnwidth,clip]{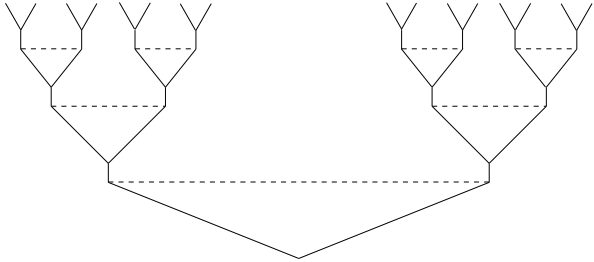}
\caption{ An example of a generalized tree-like graph.
} \label{f00}
\end{figure}

\begin{figure}
\includegraphics[width=0.6\columnwidth,clip]{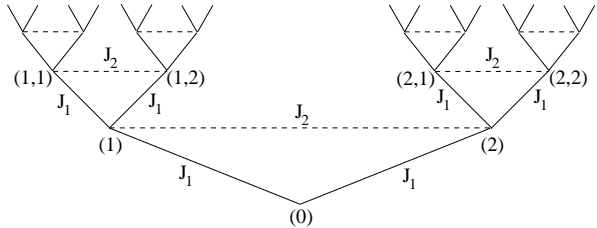}
\caption{The graph over which the competitive model
of Eq. \ref{compet} is defined: a triangular Husimi tree.
The continuous segments
constitute a Bethe lattice $\Gamma$ (an infinite tree). Dashed
segments connect vertices of $\Gamma$ belonging to the same shell.
Note that, a part the root 0, the degree of $\Gamma$ is 
$q_{\Gamma}=3$, whereas
the degree of the Husimi tree is $q_{\Gamma}'=4$.
} \label{f0}
\end{figure}

As is known, the exact solvability of a model defined over a Bethe
lattice relies on the fact that the lattice is a tree,
\textit{i.e.}, a graph without loops \cite{Baxter}. However, in
relatively recent years progresses have been made in facing,
analytically, models defined over lattices in which a finite
number of closed paths per vertex is also allowed
\cite{Lyons,Lyons2}, which we refer as ``generalized tree-like
structures'', see Fig. \ref{f00}, and models defined over graphs
in which the tree-like structure is partially broken due to the
presence of an infinite number of closed paths per vertex
as happens in 
generalized Bethe lattices (see for example \cite{SC} and references therein)
and in Husimi trees \cite{Monroe,GPW,GPW1,FarI,FarII,FarIII,FarIV}.  
The graph depicted in Fig. \ref{f0} is an example of a Husimi tree. 

Notice that, by definition, in a loop both vertices and bonds
repetitions are forbidden,
whereas in a closed path only bonds repetitions are forbidden; in a closed path
a same vertex can be crossed more than one time. 
So that, for example, in the graph of Fig. \ref{f00}, and in a regular 
$D$-dimensional lattice with $D>1$ or in the graph of Fig. \ref{f0}, 
the number of loops per vertex is finite,
but the number of closed paths per vertex is
finite only in the graph of Fig. \ref{f00}. 
This kind of difference reflects on the different
complexities in solving models defined on generalized tree-like structures
and non tree-like structures; 
whereas the first class of problems
can be exactly faced in the framework of a suitable Bethe-Peierls approach
(see for example the reviews \cite{Domb,Goltsev1} ) or,
alternatively, by using the rules given in the Refs. \cite{Lyons,Lyons2} for
determining the critical surfaces,
for the second class of problems, in general, a closed analytical method
is lacking. An exception to this is provided by models
defined over generalized Bethe lattices and Husimi trees. 
For these models in fact, the problem
of statistical mechanics can be mapped on a that of a dynamical
system involving few degrees of freedom \cite{SC,Monroe}.
In particular, the model defined over the Husimi tree of Fig. \ref{f0},
which is the subject of this paper, can be solved analytically in
a closed form.

In these kind of relatively solvable models, 
one is usually interested in studying competition
effects, so that two or more independent couplings
$J_\mathrm{1},J_\mathrm{2},\ldots$ along the links of the lattice
are given. As already investigated by numerical and
semi-analytical methods, not only on generalized Bethe lattices and Husimi
trees \cite{Monroe, V,Morita,MTA,YOS,SC,TY}, 
but even on finite dimensional lattices
\cite{E,BB}, due to the competitiveness in amplitude and sign of
the couplings and to the presence of loops giving rise to
frustration \cite{Toulouse}, a quite rich scenario of phases can take
place. In fact, besides ferro and antiferromagnetism, competing
interactions lie at the heart of a variety of original phenomena
in magnetic systems, including the existence of modulated
\cite{SC} and spin glass phases \cite{Parisi}.

After the study of the Ising and the Potts competitive models
\cite{GPW1,FarII,FarIII,Monroe} 
defined over generalized Bethe lattices and Husimi
trees with a uniform or periodic distribution of the couplings
$J_\mathrm{1},J_\mathrm{2},\ldots$, the next natural step toward
the complexity of these competitive models is the introduction of
disorder. We present here a twofold study of the
simpler Ising model with two competitive interactions
$J_\mathrm{1}$ and $J_\mathrm{2}$ defined over the triangular 
Husimi tree of Fig. \ref{f0}.
We first review this model giving the
proper interpretation of the phases in terms of the magnetization. Then we
introduce the disorder to study the triangular $\pm J$ model.
To this aim, we let
$J_\mathrm{1}$ and $J_\mathrm{2}$ to be random couplings
having a discrete probability distribution whose spread,
\textit{i.e.}, the measures of the disorder, is characterized by
two parameters, $p_1$ and $p_2$.
The analysis of the phase diagram of the random model
is done by using an effective method recently developed in \cite{MOI} and
already used in \cite{MOII}. This method consists on a mapping
between the phase diagram of the random model and the
corresponding non random one. In \cite{MOI} it has been shown
that, in the thermodynamic limit, the upper phase boundary of the
random model one finds by using this mapping, as well as the corresponding
critical behavior, become exact
whenever the graph over which the model is defined turns
out to be infinite dimensional in a broad sense, as happens for
instance in generalized tree-like structures or
generalized Bethe lattices and Husimi trees.
Unlike what emphasized in a previous version of this manuscript however,
we stress here that Bethe lattices (here meant as infinite trees) and 
Husimi trees represent a limit case where the mapping
in general is not well defined. In fact, the mapping can establish
a correspondence between the random model  
and a corresponding non random one, both built up over the same graph,  
only if their density free energies exist  
(more precisely the non
trivial part of the density free energy, see Eq. (\ref{free1})). 
We refer the reader to the recent work \cite{MOIII} for
a rigorous formulation of the mapping.
It happens unfortunately that, due to the problem of the boundary conditions
(see also the discussion in Appendix),
which in a Bethe lattice and in a Husimi tree constitute a finite fraction
of the total size of the system, 
in a Bethe lattice (here meant as an infinite tree) and in a Husimi tree, 
the density free energy,  
as a proper thermodynamic limit, does not exist
(however the circumstances for a Bethe lattice is in some sense fortunate and
the mapping turns out exact).
Therefore we advise the reader that the results we report in this
manuscript cannot be taken as exact; they have to be considered
as a mere application of the mapping being here formal and not
rigorous. Nevertheless, the equations of the mapping are still defined and,
as an approximation, they can be applied in a Husimi tree case as well. 
In fact, in studying another random version of the model, the diluted
frustrated model, in which the disorder is introduced by randomly deleting the
negative couplings $J_2$ with some probability $\mu$, 
we have found some differences between 
the results one can obtain by applying the mapping and the results one can
find by using a Bethe-Peierls-like approach \cite{Melin}. 
It turns out that, in the limit $0<-J_2\ll J_1$, this latter
method succeeds in giving the exact percolation threshold $\mu_P=1/\sqrt{2}$,
while the mapping gives $\mu_P=\sqrt{3}-1$; and a larger difference is observed
in the opposite limit $-J_2\gg J_1>0$.
Nevertheless we think that the application of the mapping also in the
Husimi tree cases provides easily important information of the random model.
We stress in fact that in the mapping no ad hoc approximation,
like the local tree-like approximation, is introduced.
 
For the present model, Eqs. (\ref{compet}-\ref{dm2}),
we have found the following scenario.
Concerning the magnetization, 
in the non random model in the frustrated regions
there is no phase transition for $|J_2|>|J_1|$ even at $T=0$, and this
is physically explained by looking directly at the ground state
that becomes equivalent to a gas of non interacting dimers with
coupling $J_2$ (also called a spin liquid in the literature).
In the $\pm J$ triangular model instead, even for $p_1=p_2=1$,
there exists always a stable spin glass transition and,
for suitable values of $p_1$ and $p_2$ a ferro or antiferromagnetic
disordered transition may exist as well. Whereas the latter transition
can be read just as due to an effective renormalization of the
couplings $J_1$ and $J_2$, the former transition is interpreted as
due to the onset of the overlap ordering of the dimers, as signaled
by the study of the exact ground states of the non random model which is
equivalent to the random one for $p_1=p_2=1$.
Furthermore, we observe that in the limit $|J_2|\gg |J_1|$ the picture
in terms of non interacting dimers is recovered also in the
glassy phase. 

We shall limit ourself mainly to the study of the critical
surfaces inferring little on the nature of the several phases.
Furthermore, we recall that for negative couplings,
the correspondence between the solution of the Ising model on the Bethe
lattice (the graph of Fig. \ref{f0} with $J_2=0$)
and that on a corresponding sparse system (like the Erd$\mathrm{\ddot{o}}$s 
R$\mathrm{\acute{e}}$nyi random graph), 
where long global frustrated loops are
also present, is lost.
Therefore, following the same philosophy of, \textit{e.g.},
the Refs. \cite{V,Morita,SC},
the solution we will present here concerns,
strictly speaking, only an Ising model model defined over a structure where global
loops are absent.
Nevertheless, we believe that since we take into account
the effect of all the other explicit loops, at least for certain 
aspects, the solution we find
sheds some light also on a possible sparse representation of the model
even in the frustrated regions. In fact, the ground state we describe
in Sec. 3 has a quite strong similarity to that one finds in the anisotropic 
triangular Ising model \cite{Likos}.

The competitive non random model was partially already analyzed in
\cite{GPW1,FarII,Melin}. Notice that it is different from the model considered
in the Ref. \cite{Monroe} (third item); 
in fact, in the non random case, its relative simplicity allows 
an exact analytical study.
Note also that in the Refs. \cite{GPW1,FarII,Melin} 
no detail about the ground states was given.

Very recently, a renewed attention has been attracted by 
the role of loops both in
statistical mechanics and computer science
\cite{Ginestra,MR,PS,MS}. In fact, even if the most used
Bethe-Peierls approximation usually works well in many important
models near to be tree-like, rigorously speaking, it becomes exact
only in perfect tree-structures and generalized tree-like structures, like
the one depicted in Fig. \ref{f00}, whereas in more complex graphs like
that of Fig. \ref{f0}, can be wrong.
Successful efforts have been made
toward a general treatment of the problem and systematic
corrections to the Bethe-Peierls approximation have been now
rigorously established \cite{CC}. 
However, as a general rule, we recall that between the complexity,
both analytical or numerical, in solving a random model and a corresponding
non random one there is a gap. In fact, in general the approach used to
solve, possibly exactly, the latter, does not lend itself to be simply
adapted to the former.
Our study, even if limited to the knowledge of the upper
critical surface and to the critical behavior, 
constitutes an example of a general procedure which, at least in the cases
where the conditions for the mapping are satisfied, covers this gap.

The paper is organized as follows. In Sec. 2
we introduce the uniform and the random model defined over
the triangular Husimi tree of Fig. \ref{f0}.
In Sec. 3 we analyze the phase diagram of the uniform model
giving the proper interpretation of the invariant ground states.
In Sec. 4 we recall the mapping through which the phase diagram of
the random model will be analyzed in the following Sec. 5.
Finally, some conclusions are drawn in Sec. 4.
The Appendix A is devoted to Sec. 3.

\section{Competitive Ising models on a structure with loops}
\label{models}
Let us consider a semi-infinite Bethe lattice $\Gamma$ in which any
vertex, but a singular one 0, has coordination number $q_{\Gamma}=3$ (\textit{i.e.}
branching number $k=2$) as in Fig. \ref{f0}. Chosen $0$ as a root
vertex, it is convenient to divide $\Gamma$ in shells
$S_1,S_2,\ldots,S_n,\ldots$ \cite{Baxter}. Given a couple of
vertices $(x,y)$ of $\Gamma$, we will write $(x,y)$ as $<x,y>$
if $x$ and $y$ are nearest neighbors, \textit{i.e.}, 
if they are connected through a bond of $\Gamma$,
and $(x,y)$ as $>x,y<$ if $x$ and
$y$ belong to the same shall $S_n$ and are at distance $d(x,y)=2$ 
on $\Gamma$.
\subsection{Non random case}
Let us consider the following Ising model with competing interactions
\begin{eqnarray}
\label{compet}
H(\s)=-J_\mathrm{1}\sum\limits_{<x,y>}{\s(x)\s(y)}-J_\mathrm{2}
\sum\limits_{>x,y<}{\s(x)\s(y)}
\end{eqnarray}
where $J_\mathrm{1},J_\mathrm{2}\in \mathop{\mathrm{R}}$
are coupling constants and $\s$ is a configuration of spin variables
$\s_x=\pm 1$ over the vertices of $\Gamma$.
Far from the root the graph over which this Ising model is defined,
is characterized by the fact that at any vertex is present
a triangular loop with two couplings $J_\mathrm{1}$ and one coupling
$J_\mathrm{2}$
as indicated in Fig. \ref{f0}. For $J_2\neq 0$ the model is
defined over the Husimi tree of Fig. \ref{f0} having degree $q_{\Gamma}'=4$.

\subsection{Random case (spin glass)}
In the random version of the competitive model of Eq.
(\ref{compet}), the couplings $J_\mathrm{1}$ and $J_\mathrm{2}$ become
independent random variables so that the corresponding Hamiltonian
for the quenched system is
\begin{eqnarray}
\label{Rcompet}
H(\s)=-\sum\limits_{<x,y>}J_{x,y}^{(1)}{\s(x)\s(y)}-
\sum\limits_{>x,y<}J_{x,y}^{(2)}{\s(x)\s(y)}.
\end{eqnarray}
The free energy $F$ is defined by
\small
\begin{eqnarray}
\label{logZ}
-\beta F\equiv \int d\mathcal{P}\left(\{J_{x,y}^{(1)}\},
\{J_{x,y}^{(2)}\}\right)
\log\left(Z\left(\{J_{x,y}^{(1)}\},\{J_{x,y}^{(2)}\}\right)\right),
\end{eqnarray}
\normalsize where
$Z\left(\{J_{x,y}^{(1)}\},\{J_{x,y}^{(2)}\}\right)$ is the
partition function of the quenched system
and $d\mathcal{P}\left(\{J_{x,y}^{(1)}\},\{J_{x,y}^{(2)}\}\right)$
is a product measure given by \small
\begin{eqnarray}
\label{dP}
d\mathcal{P}\left(\{J_{x,y}^{(1)}\},\{J_{x,y}^{(2)}\}\right)=
\prod_{<x,y>} d\mu_\mathrm{1}\left( J_{x,y}^{(1)} \right)
\prod_{>x,y<} d\mu_\mathrm{2}\left( J_{x,y}^{(2)} \right),
\end{eqnarray}
\normalsize $d\mu_\mathrm{1}$ and $d\mu_\mathrm{2}$ being two
given normalized measures featuring the disorder.
In general this model, besides a disordered
ferromagnetic or antiferromagnetic transition (P-F/AF), may
manifest a spin glass transition ({P-SG}). A generic inverse
critical temperature of the random model will be indicated by
$\beta_c$. We will consider the following choice: 
\subsubsection{Triangular $\pm J$ model}
\small
\begin{eqnarray}
\label{dm1}
d\mu_\mathrm{1}\left( J_{x,y}^{(1)} \right)
&=&p_\mathrm{1}\delta(J_{x,y}^{(1)}
-J_\mathrm{1})+(1-p_\mathrm{1})\delta(J_{x,y}^{(1)}+J_\mathrm{1}), \\
\label{dm2}
d\mu_\mathrm{2}\left( J_{x,y}^{(2)} \right)
&=&p_\mathrm{2}\delta(J_{x,y}^{(2)}-J_\mathrm{2})+
(1-p_\mathrm{2})\delta(J_{x,y}^{(2)}+J_\mathrm{2}),
\end{eqnarray} 
\normalsize
where $J_\mathrm{1}$ and $J_\mathrm{2}$, as in the non random
model, are arbitrary parameters and $p_\mathrm{1},p_\mathrm{2}\in
[0,1]$ characterize the disorder, maximum at
$p_\mathrm{1}=p_\mathrm{2}=0.5$ and minimum for $p_\mathrm{1}$ and
$p_\mathrm{2}$ equal to 0 or 1.

\section{Analysis of the non random model:
phase diagram - gas of dimers}
\begin{figure}
\begin{center}
\includegraphics[width=0.6\columnwidth,clip]{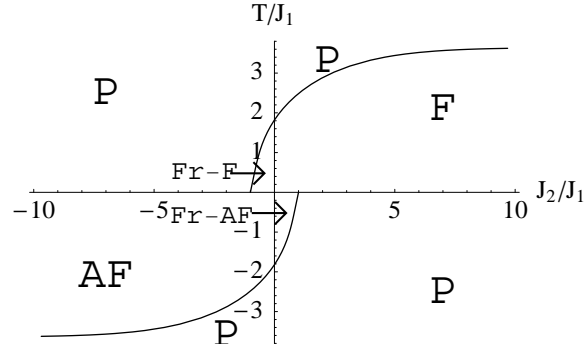}
\end{center}
\caption{ Phase diagram of the competitive model, Eq. (\ref{compet}).
The transition lines {P-F} and {P-AF} come from the solution
of Eqs. (\ref{F}) and (\ref{AF}), respectively. {Fr-F} and {Fr-A}F stand
for the two frustrated ferro and antiferromagnetic regions 
with $J_1>0$ and $J_2<0$, and
$J_1<0$ and $J_2<0$, respectively. At T=0, the {P-F} and {P-AF} lines
cross the $J_2/J_1$ axis at $J_2/J1=-1$ and $J_2/J_1=1$, respectively.
}
\label{f1}
\end{figure}

\begin{figure}
\begin{center}
\includegraphics[width=0.6\columnwidth,clip]{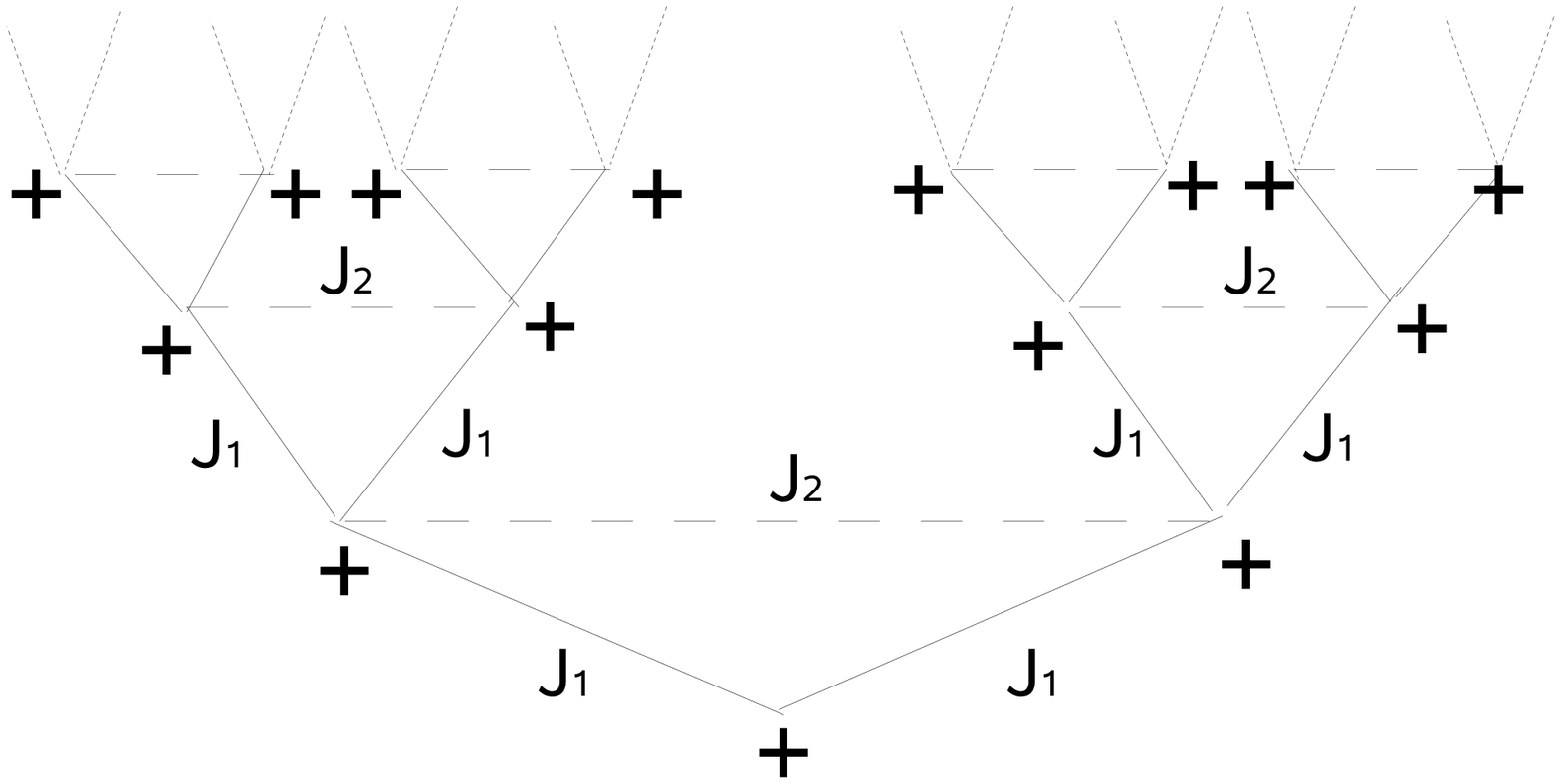}
\end{center}
\caption{One of the two ground states
of the competitive model, Eq. (\ref{compet}), in the Fr-F region
($J_1>0$, $J_2<0$ and $J_1>|J_2|$).
}
\label{gs}
\end{figure}

\begin{figure}
\begin{center}
\includegraphics[width=0.6\columnwidth,clip]{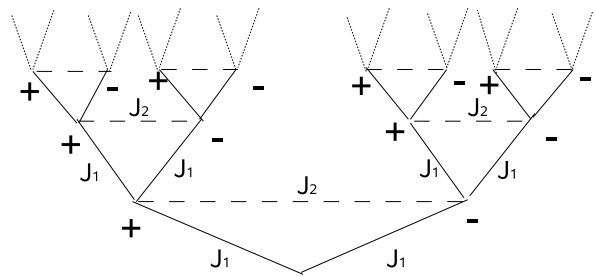}
\end{center}
\caption{One of the infinitely many degenerate ground states
of the competitive model, Eq. (\ref{compet}), in the regions
$J_1>0$, $J_2<0$ or $J_1<0$, $J_2<0$ with $|J_1|>|J_2|$.
This phase can be read as a gas of non interacting
dimers with antiparallel spins. A dimer here is defined as constituted
by two spins belonging to the same shall and connected through a bond
with coupling $J_2$.
}
\label{gs_2}
\end{figure}

\begin{figure}
\begin{center}
\includegraphics[width=0.6\columnwidth,clip]{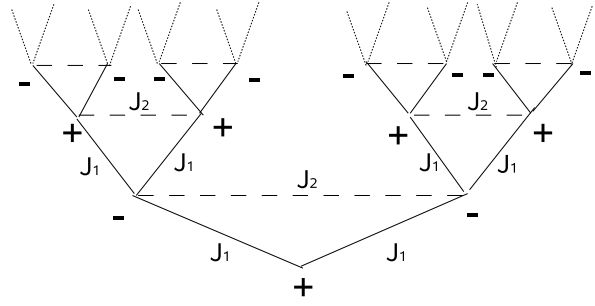}
\end{center}
\caption{One of the two ground states
of the competitive model, Eq. (\ref{compet}), in the Fr-AF region
($J_1<0$, $J_2<0$ and $|J_1|>|J_2|$).
}
\label{gs_1}
\end{figure}

Concerning the phase transitions,
the model defined in Sec. 2.1, even in the presence of an 
external magnetic field $h$, has been recently solved analytically
in \cite{GPW1}. In Appendix A we report another approach for deriving
the phase transitions. There are two critical lines given by
\begin{eqnarray}
\label{F}
\t_\mathrm{2}=\frac{2\t_\mathrm{1}}{\t_\mathrm{1}^2-3},
\quad \t_\mathrm{1}>\sqrt{3},
\end{eqnarray}
\begin{eqnarray}
\label{AF}
\theta_\mathrm{2}=\frac{2\theta_\mathrm{1}}{1-3\theta^2_\mathrm{1}},
\quad \theta_\mathrm{1}<1/\sqrt{3},
\end{eqnarray}
where the parameters $\theta_\mathrm{1}$ and $\t_\mathrm{2}$ are defined as
\begin{eqnarray}
\theta_\mathrm{1}=e^{2\beta J_\mathrm{1}}, \quad \t_\mathrm{2}=
e^{2\beta J_\mathrm{2}}.
\end{eqnarray}
In Fig. \ref{f1} we show the corresponding phase diagrams obtained
as solutions of Eqs. (\ref{F}) and (\ref{AF}) along
the axis $T/J_\mathrm{1}$ and $J_\mathrm{2}/J_\mathrm{1}$.
Equations (\ref{F}) and (\ref{AF}) describe, respectively, a
paramagnetic-ferromagnetic ({P-F}) transition, in which $J_1>0$,
and a paramagnetic-antiferromagnetic (P-AF) transition in which $J_1<0$.
By a direct inspection of the minimum of $H$, Eq. (\ref{compet}), 
it follows that, in the two non paramagnetic frustrated ({Fr}) regions
$J_1>0,J_2<0$ and $J_1<0,J_2<0$, see Fig. \ref{f1},
the corresponding ground states, despite the frustration, are simply {F} and
{AF}, respectively. The reason for that lies on the smallness
(in modulo) of the coupling $J_2$ in the two mentioned regions.
More precisely, by looking at the configurations minimizing the energy,
it is easy to see that, if \textit{e.g.} $J_1>0$ and $J_2<0$, for
$J_1>|J_2|$ there exist only two ground states obtained by taking all the
spins parallel, the corresponding ground state energy per triangle
being $e=-2J_1-J_2$ (see Fig. \ref{gs}); 
whereas for $J_1<|J_2|$ the ground state
becomes infinitely degenerate, the corresponding ground
state energy per triangle being $e'=J_2$ (see Fig. \ref{gs_2}).
In the latter case the whole set of ground
states is obtained by taking in all the possible ways the spins
connected by the couplings $J_2$ (\textit{i.e.}, the spins on the same shell
at distance $d=2$ on $\Gamma$) as antiparralel.
A similar conclusion holds also in the other frustrated
region where $J_1,J_2<0$. In this case, when $|J_1|>|J_2|$,
the ground state is obtained by taking alternated spins on the
Bethe lattice $\Gamma$ (note that, as a consequence, the spins are not
alternated on the Husimi tree; spins on the same shall
are parallel, see Fig. \ref{gs_1}); 
whereas for $|J_1|<|J_2|$ the ground state has the same
structure of the case $J_1>0$ and $|J_1|<|J_2|$ above analyzed 
(Fig. \ref{gs_2}).
In other words, for $T=0$ in the {Fr} regions, crossing the critical
{P-F} or {P-AF} lines amounts to move oneself between a connected set
of triangles with ordered or antiordered spins (on $\Gamma$), respectively,
and a set of independent
dimers having at their edges two antiparallel spins
interacting with a negative coupling $J_2$.
In the sectors II and IV of Fig. \ref{f1}, at low temperature,
the {P} phase corresponds
therefore to a gas of non interacting dimers
having each one two possible degenerate states.
Note that, as a consequence, in the regions $|J_2|>|J_1|$, 
for any selected ground state $\alpha$,
the spatial average magnetization $m_\alpha$ is zero, 
but the Edwards Anderson order parameter $q_{EA}=q_{\alpha,\alpha}$ is 1,
so that, in certain aspects, 
the zero temperature limit in these regions leads 
to a sort of glassy phase at zero temperature~\cite{Nota}.
We note that, in this phase, the interesting
``objects'' are not the spins, but the dimers. 
In fact, by looking at the moments of $P(q)$, the probability distribution
of the overlaps between spins \cite{Parisi}, and using the symmetry
among the ground states one has $P(q)=\delta(q)$, whereas for the overlap
between dimers one has $P^D(q^D)=\delta(q^D-1)$, where for overlap between
dimers $q^D$ we mean 
$q_{\alpha,\beta}^D\equiv 
1/N\sum_i<\sigma_i^x\sigma_i^y>_\alpha<\sigma_i^x\sigma_i^y>_\beta$,
where $\alpha$ and $\beta$, with $\alpha\neq\beta$, 
label two ground states, the index $i$ stands for a dimer's index and 
the superscripts $x$ and $y$ label the two spins of a dimer. 

In the literature 
this phase was called a spin liquid phase \cite{Melin}.
An important question to be addressed is whether 
this zero temperature phase signals the existence
of a spin glass transition at some finite temperature or
it is merely a feature of the zero temperature limit.
In the next sections by applying the mapping we will see that  
the former hypothesis emerges naturally.  


\section{Mapping}
\subsection{Equations of the mapping}
Let 
\begin{eqnarray}
\label{uv}
t_1&=&\tanh(\beta J_\mathrm{1}), \\
t_2&=&\tanh(\beta J_\mathrm{2}).
\end{eqnarray}
Later on, it will be useful to consider the following decomposition
of density free energy
$f_I$ of the non random model in the {P} phase
(when necessary we will use the suffix $I$
for stressing that the given quantity refers to the non random
Ising model, as opposed to the random one)
\small
\begin{eqnarray}\label{free1}
-\beta f_I &=&\log 2 +\log(\cosh(\beta J_\mathrm{1}))
+\frac{1}{2}\log(\cosh(\beta J_\mathrm{2}))+ \varphi_I(t_1,t_2).
\end{eqnarray} \normalsize
The term $\varphi_I$ represents the non trivial part of the high temperature
expansion of the density free energy, responsible for the critical
behavior of the system.

By using a recently proposed method \cite{MOI,MOII,MOIII}, we will find
the phase boundary of the random model by mapping it to
the non random one. According to this mapping, to
find the upper phase boundary of the random model we consider the
transition equations (\ref{F}) and (\ref{AF}) of the non random
model and perform one of the two following substitutions
\begin{eqnarray}
\label{Fmapp}
\left\{
\begin{array}{l}
t_1=\tanh(\beta J_1) \rightarrow
t_1^{(\mathrm{F})}\equiv
\int d\mu_\mathrm{1}\left(J_{x,y}^{(1)}\right)
\tanh\left(\beta J_{x,y}^{(1)}\right) \\
t_2=\tanh(\beta J_2) \rightarrow
t_2^{(\mathrm{F})}\equiv
\int d\mu_\mathrm{2}\left(J_{x,y}^{(2)}\right)
\tanh\left(\beta J_{x,y}^{(2)}\right),
\end{array}
\right.
\end{eqnarray}
\begin{eqnarray}
\label{SGmapp}
\left\{
\begin{array}{l}
t_1=\tanh(\beta J_1) \rightarrow
t_1^{(\mathrm{SG})}\equiv \int d\mu_\mathrm{1}
\left( J_{x,y}^{(1)} \right)\tanh^2\left(\beta J_{x,y}^{(1)}\right) \\
t_2=\tanh(\beta J_2) \rightarrow
t_2^{(\mathrm{SG})}\equiv
\int d\mu_\mathrm{2}\left( J_{x,y}^{(2)} \right)
\tanh^2\left(\beta J_{x,y}^{(2)}\right).
\end{array}
\right.
\end{eqnarray}
The substitution (\ref{Fmapp}) in Eqs. (\ref{F}/\ref{AF}) provides
the disordered ferro/antiferromagnetic transitions (P-F/AF),
whereas the substitution (\ref{SGmapp}) in Eqs. (\ref{F}) provides
the spin glass transition ({P-SG}). For a given couple of
parameters $(J_\mathrm{1},J_\mathrm{2})$, we find 
the corresponding critical inverse
temperatures $\beta_c^{(F/AF)}$ or $\beta_c^{(\mathrm{SG})}$, respectively.
In the thermodynamic limit, for given parameters $(J_\mathrm{1},J_\mathrm{2})$,
between the two critical values $\beta_c^{(F/AF)}$ and
$\beta_c^{(\mathrm{SG})}$, only one is stable and the system is critical at
$\beta_c$ given by
\begin{eqnarray}
\label{beta}
\beta_c=\mathrm{min}\{\beta_c^{(\mathrm{SG})},\beta_c^{(F/AF)}\}.
\end{eqnarray}

By performing the substitutions (\ref{Fmapp}) or (\ref{SGmapp}), 
the transition equations
(\ref{F}) and (\ref{AF}) take respectively the form
\begin{eqnarray}
\label{Fmapp1} t_2^{(\s)}((t_1^{(\s)})^2-2t_1^{(\s)})+1-2t_1^{(\s)}=0,
\end{eqnarray}
\begin{eqnarray}
\label{AFmapp1} t_2^{(\s)}((t_1^{(\s)})^2+2t_1^{(\s)})+2t_1^{(\s)}+1=0,
\end{eqnarray}
where $\s=F$ for the substitution (\ref{Fmapp}) and
$\s=SG$ for the substitution (\ref{SGmapp}).

The multicritical point P-F-SG, if any, 
is solution of the system of equations obtained
by taking the two versions of Eq. (\ref{Fmapp1}) with the two
substitutions (\ref{PF}) and (\ref{PSG}). Similarly for the
multicritical point P-AF-SG from the two versions of Eq. (\ref{AFmapp1}).

\subsection{Critical behavior}
The transformations (\ref{Fmapp}) and (\ref{SGmapp}) can also be used
to calculate the free energy and the correlation functions infinitely near
the upper phase boundary above the {P-F} and the {P-SG} lines, respectively.
To this aim, besides $\varphi_I$, we define also $\varphi$,
that is, the non trivial part of the high temperature expansion of the
density free energy $f$ of the random model:
\begin{eqnarray}\fl
\label{free}
-\beta f&=&\log 2 +\int d\mu_\mathrm{1}\log(\cosh(\beta J_\mathrm{1}))+
\frac{1}{2}\int d\mu_\mathrm{2}\log(\cosh(\beta J_\mathrm{2}))+\varphi.
\end{eqnarray}
According to \cite{MOI}, in the {P} phase, infinitely near the
{P-F} and the {P-SG} lines, $\varphi$ takes respectively the two
following forms:
\begin{eqnarray}
\label{freePF}
\varphi_I(t_1,t_2) \rightarrow \varphi^{(\mathrm{F})}&=&
\varphi_I(t_1^{(\mathrm{F})},t_2^{(\mathrm{F})}), \\
\label{freePSG}
\varphi_I(t_1,t_2) \rightarrow \varphi^{(\mathrm{SG})}&=&\frac{1}{2}
\varphi_I(t_1^{(\mathrm{SG})},t_2^{(\mathrm{SG})}).
\end{eqnarray}

Similar formulae hold for the correlation functions at zero external field.
In other words, we get the exact critical behavior of the system in the
upper critical line in the {P} region. On the other hand, as a general rule of the
mapping, we immediately see that the critical exponents of the
random system cannot be affected by the substitutions
(\ref{Fmapp}) or (\ref{SGmapp}), so that the critical behavior of
the two systems in the {P} region is simply the same.
Therefore, since the non random models defined over graphs of interest, 
like generalized Bethe lattices and Husimi tree (and many random graphs), 
have a mean field-like behavior with a second order phase transition, 
the same happens for the corresponding random versions.

\subsection{Condition for the mapping - 
the case of Bethe lattices and Husimi trees}
\begin{figure}
\includegraphics[width=0.6\columnwidth,clip]{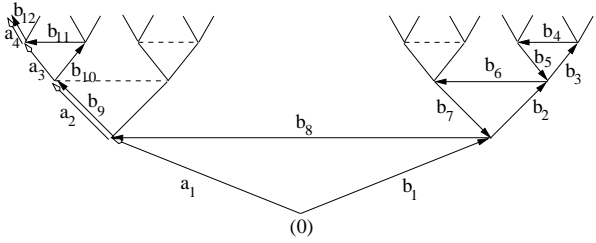}
\caption{Two paths starting from the vertex 0 are drawn. The bonds
of the two paths are indicated as $a_1,a_2,\ldots$ and
$b_1,b_2,\ldots$, respectively. The two paths overlap each other
in correspondence of the bonds $a_2=b_9$ and $a_4=b_{12}$. These
bonds on the figure are slightly shifted for visual convenience. }
\label{f}
\end{figure}

In \cite{MOI} it has been shown that the upper critical boundary
one obtains by using this mapping is exact whenever the
underlying set of links is infinite dimensional in a broad sense
as happens for instance on generalized tree-like structures
having a finite number of loops per vertex.
The condition for the mapping to be exact (infinite
dimensional in the broad sense) is that $\varphi_I$ exists (see Sec. I) 
and that choosing randomly two
arbitrary infinite long paths passing through a vertex, the
probability $p(l)$ that they overlap each other for $l$
bonds goes to zero exponentially as $l$ goes to infinity \cite{Nota1}.
A path of length $l$ is defined
as a succession of $l$ different bonds connecting vertices.
In the framework of the high temperature expansion of the partition function, 
one sees that near the critical temperature, infinitely long closed
paths correspond to the contributions of the free
energy, while open paths and combinations of open and closed paths
correspond to the contributions of the correlation functions.
The mentioned probability for the overlapping of two
paths is a feature that applies both to closed and open paths so
that, for calculating $p(l)$ we are free to consider both 
open or closed paths.
We can see easily that for our model $p(l)\to 0$
exponentially as follows.
Given $l$, let us consider two arbitrary paths of lengths
$l_1$ and $l_2$ with $l_1,l_2\geq l$. Let us refer to these paths
as first and second path, respectively. Since the paths are all
statistically equivalent, we can choose for simplicity the first
path to coincide with the straight line indicated in Fig. \ref{f}
(located in the most left part of the figure). For the second path
to overlap $l$ times with $l$ bonds of the first path, the second
path must cross $l$ vertices belonging to the first path. On the
other hand, at any of such a crossing point, the second path has
two over four possibilities (2/4) to make an overlap with a bond
of the first path and, if the number of crossing points has to be
infinite, as it must be for the requirement $l\to\infty$, it is
easy to see that only one of the above possibilities can be taken
(choosing the bond going in the upward direction). Therefore, we
see that $p(l)<1/2^l$ and for $l$ large we have $p(l)\sim 1/4^l$,
\textit{i.e.}, an exponential decay. 
In general the same conclusion
holds for any non trivial (that is with a vertex degree $q>2$) generalized
Bethe lattice and Husimi tree. Note that, in general, 
in such structures the
number of closed paths passing through a vertex is infinite 
(see Sec. 1).

\section{Phase diagram of the random model}
In the two following subsections we apply the mapping to study the
phase diagram of the disordered model introduced in Sec. 2.2. 
Note that, as manifested in Fig. \ref{f1}, the {P-F} and the {P-AF}
transitions are completely asymmetric one another and since they
can occur only for positive and negative values of $J_\mathrm{1}$,
respectively, it will be sufficient to consider only the {P-F} and
the {P-SG} transitions. Therefore, from now on we will study only
the region with $J_\mathrm{1}\geq0$ 
corresponding to Eqs. (\ref{F}) and (\ref{Fmapp1}) that here we rewrite
\begin{eqnarray}
\label{Fmapp1bis} t_2^{(\s)}((t_1^{(\s)})^2-2t_1^{(\s)})+1-2t_1^{(\s)}=0, 
\quad \sigma=\mathrm{F,SG}.
\end{eqnarray}
We observe that Eq. (\ref{Fmapp1bis}) (as well as Eq. (\ref{AFmapp1})), 
with respect to $t_1^{(\sigma)}$, are quadratic and biquadratic under the
transformations (\ref{PF}) and (\ref{PSG}), respectively, so that
we can solve this equation explicitly with respect to $t_1^{(\sigma)}$. This
procedure turns out to be convenient for studying some asymptotes since we want to
plot the critical lines on the plane $(T/J_1,J_2/J_1)$. 
Let us put
\begin{eqnarray}
\kappa(x)\equiv\frac{1+x-\sqrt{1+x+x^2}}{x}.
\end{eqnarray}
In terms of $t_1^{(\sigma)}$, the solution of Eq. (\ref{Fmapp1bis}) 
for $t_2^{(\sigma)}$
can be expressed as
\small
\begin{eqnarray}
\label{general}
t_1^{(\sigma)}=\kappa(t_2^{(\sigma)}).
\end{eqnarray} \normalsize

\subsection{Triangular $\pm J$ model}
For the triangular $\pm J$ model the randomness is given through
Eqs. (\ref{dm1}) and (\ref{dm2}) which, plugged in the
transformations (\ref{Fmapp}) and (\ref{SGmapp}), give respectively
\begin{eqnarray}
\label{PF}
\left\{
\begin{array}{l}
t_1^{(\mathrm{F})}=(2p_\mathrm{1}-1)t_1,\\
t_2^{(\mathrm{F})}=(2p_\mathrm{2}-1)t_2,
\end{array}
\right.
\end{eqnarray}
and
\begin{eqnarray}
\label{PSG} \left\{
\begin{array}{l}
t_1^{(\mathrm{SG})}=t_1^2,\\
t_2^{(\mathrm{SG})}=t_2^2.
\end{array}
\right.
\end{eqnarray}
Note that in the last case there is no dependence on the values on
$p_\mathrm{1}$ and $p_\mathrm{2}$, so that in the triangular $\pm J$ model 
model the {P-SG} transition is unique. Note also that  Eqs. (\ref{PF}) are
unchanged under the exchanges $p_\mathrm{1}\to 1-p_\mathrm{1}$ and
$J_\mathrm{1}\to -J_\mathrm{1}$, and $p_\mathrm{2}\to
1-p_\mathrm{2}$ and $J_\mathrm{2}\to -J_\mathrm{2}$, respectively.
Therefore, it will be sufficient to restrict ourself to the region
\begin{eqnarray}
\label{pp}
\left\{
\begin{array}{l}
p_\mathrm{1}\geq 0.5, \\
p_\mathrm{2}\geq 0.5.
\end{array}
\right.
\end{eqnarray}

From (\ref{general}) the solution for the {P-SG} line (in terms of $t_2^{(\mathrm{SG})}$) is
\small
\begin{eqnarray}
\label{PSGG}
\frac{T_c^{(\mathrm{SG})}(J_\mathrm{2})}{J_\mathrm{1}}=\left[\Arctanh\left(\kappa(t_2^{(\mathrm{SG})})
\right)^{1/2}\right]^{-1},
\end{eqnarray} \normalsize

while the solution for the {P-F} lines (in terms of $t_2^{(\mathrm{F})}$) is \small
\begin{eqnarray}
\label{PFG}
\frac{T_c^{(\mathrm{F})}(J_\mathrm{2})}{J_\mathrm{1}}=
\left[\Arctanh\left(\frac{\kappa(t_2^{(\mathrm{F})})}{2p_\mathrm{1}-1}\right)\right]^{-1}.
\end{eqnarray} \normalsize

As an argument based on the high temperature expansion
of the free energy suggests, we see in Fig. \ref{f2} that the presence of loops,
tuned by fixing $J_\mathrm{1}$ and increasing the amplitude of the parameter
$J_\mathrm{2}$, increases the spin glass critical temperature $T_c^{(\mathrm{SG})}$.
In particular, the values of $T_c^{(\mathrm{SG})}$ for $J_\mathrm{2}\to 0$ and
$J_\mathrm{2}\to\infty$ are given respectively by \small
\begin{eqnarray}
\label{PSG0}
\frac{T_c^{(\mathrm{SG})}(J_\mathrm{2}\to 0)}{J_\mathrm{1}}=
\left[\Arctanh\left(\frac{1}{\sqrt{2}}\right)\right]^{-1}=1.13459,
\end{eqnarray}
\begin{eqnarray}
\label{PSGinfty}
\frac{T_c^{(\mathrm{SG})}(J_\mathrm{2}\to\infty)}{J_\mathrm{1}}=
\left[\Arctanh\left(\sqrt{2-\sqrt{3}}\right)\right]^{-1}=1.74487.
\end{eqnarray} \normalsize
As we shall see below, for values of $p_\mathrm{1}$ and
$p_\mathrm{2}$ sufficiently high and far from $0.5$ and for
$J_\mathrm{2}>0$, a similar behavior holds for the disordered
ferromagnetic transition as well. Note however that unlike Eq.
(\ref{PSGG}), for Eq. (\ref{PFG}) exist regions without solution.
Finally, observe that, for $J_\mathrm{1}$ finite, the asymptotic
values of $T_c^{(\mathrm{SG})}$ or $T_c^{(\mathrm{F})}$ for $J_\mathrm{2}=0$ and
$J_\mathrm{2}=+\infty$ correspond to the values of a pure Bethe
lattice case and to a system with an infinitely strong loops
effects, respectively. Note that in the latter case, if
$J_\mathrm{1}\to 0$, $T_c^{(\mathrm{SG})}$ and $T_c^{(\mathrm{F})}$ 
also $\to 0$, the
lattice being just an infinite set of non interacting dimers of spins:
a gas of dimers. We point out that this last situation differs from the one
in which $J_2\to \pm\infty$ and $J_1$ remains finite: an arbitrary small but
non zero coupling $J_1$ make the structure with triangular loops non reducible
to dimers.  

\subsubsection{The case $p_\mathrm{1}=p_\mathrm{2}$.}
\begin{figure}
\begin{center}
\includegraphics[width=0.6\columnwidth,clip]{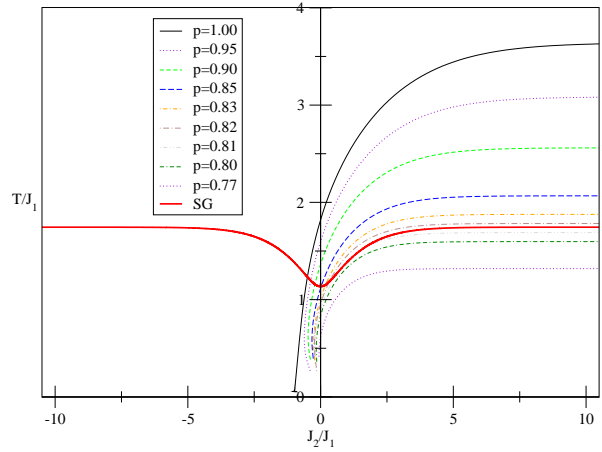}
\end{center}
\caption{Critical lines of the disordered transition {P-F} for several
values of $p\equiv p_\mathrm{1}=p_\mathrm{2}$ and the {P-SG} line. Note that,
according to Eq. (\ref{beta}), for any given $p$,
there is a unique stable upper critical line
obtained by taking the convex construction as shown in Fig. \ref{f4}.
}
\vspace{0.5cm}
\label{f2}
\end{figure}

Here we discuss an analysis carried out for the simpler case
$p\equiv p_\mathrm{1}=p_\mathrm{2}$.
We have performed a study for the disordered {P-F} transition for
several values of $p$. In Fig. \ref{f2} we report these lines.
For $p=1$ one recovers the non random ferromagnetic transition
(Fig. \ref{f1}).
In general for fixed parameters $J_\mathrm{1}$ and $J_\mathrm{2}$,
as the value of $p$ is decreased from 1.00 toward 0.50,
the critical temperature $T_c^{(\mathrm{F})}$ decreases continuously.
For a fixed value of $p$, as the value of $J_\mathrm{2}$ increases the {P-F} lines
reach a horizontal asymptote whose value can be obtained explicitly from
Eq. (\ref{PFG})
\begin{eqnarray}
\label{Jinfty} \frac{T_c^{(\mathrm{F})}(J_\mathrm{2}\to
+\infty)}{J_\mathrm{1}}=
\left[\Arctanh\left(\frac{\kappa(A)}{A}\right) \right]^{-1},
\end{eqnarray}
where $A=2p-1$. When $J_2$ decreases toward 0 from Eq. (\ref{PFG}) we get
\begin{eqnarray}
\label{PSG0}
\frac{T_c^{(\mathrm{F})}(J_\mathrm{2}\to 0)}{J_\mathrm{1}}=
\left[\Arctanh\left(\frac{1}{2A}\right)\right]^{-1}.
\end{eqnarray}
Finally for negative values of $J_2$ the {P-F} lines
reach a final point given by Eq. (\ref{PFG})
\begin{eqnarray}
\label{PSGneg} \frac{T_c^{(\mathrm{F})}(J_\mathrm{2}\to
-\infty)}{J_\mathrm{1}}=
\left[\Arctanh\left(\frac{\kappa(-A)}{A}\right)\right]^{-1}.
\end{eqnarray}
Note that, for fixed $J_1$, Eq. (\ref{Jinfty}) has no solution
for values of $p\leq 0.6909$, whose asymptote corresponds
to $T_c^{(\mathrm{F})}/J_\mathrm{1}=0$. Similarly, for fixed $J_1$ and
for $J_\mathrm{2}\to -\infty$ Eq. (\ref{PSGneg}) has
solution only for the pure ferromagnetic case $p=1$.
Note however that
in both the cases the above temperatures $T_c^{(\mathrm{F})}$
are lower than $T_c^{(\mathrm{SG})}$ so that, according to
Eq. (\ref{beta}), the actual phase in these regions is {SG} and the above
$T_c^{(\mathrm{F})}$'s represent only unstable (not physically relevant) transitions.

\begin{figure}
\begin{center}
\includegraphics[width=0.6\columnwidth,clip]{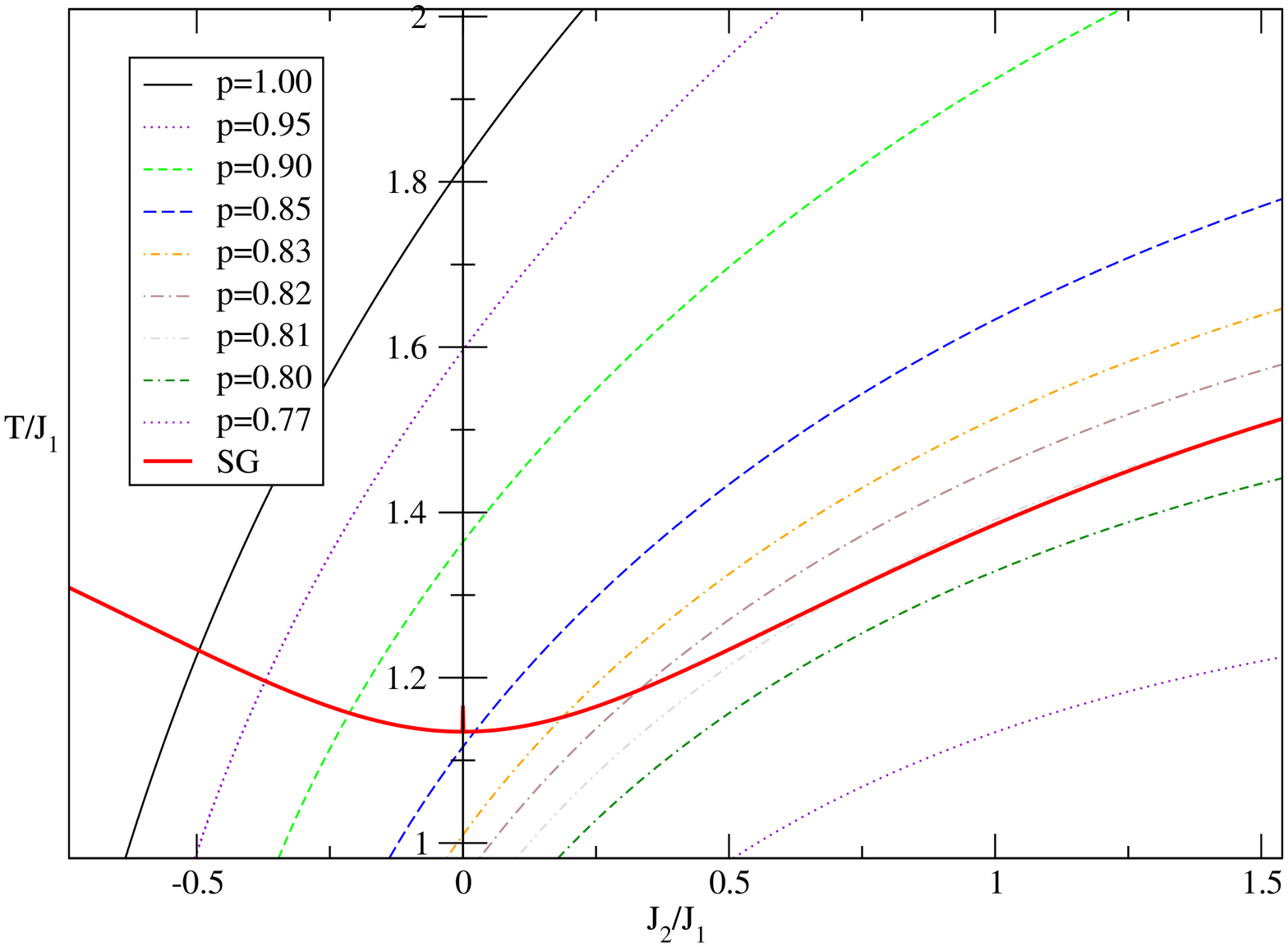}
\end{center}
\caption{ Enlargement of Fig. \ref{f2} around the {SG} line.
}
\label{f3}
\end{figure}

By using Eq. (\ref{beta}), in the plane
$(T/J_\mathrm{1},J_\mathrm{2}/J_\mathrm{1})$,
as evident from Figs. \ref{f2} and \ref{f3}, one finds the following scenario.
Let us consider first the case $J_\mathrm{2}\geq 0$. For $p\geq 0.85
\pm 0.01$ for any value of the parameters $J_\mathrm{1}$ and $J_\mathrm{2}$ the
transition can be only {P-F}. For $p\in [0.81,0.85]\pm 0.01$
there is a range of values $J_\mathrm{1}$ and $J_\mathrm{2}$ where the stable
transition is {P-SG} and a range where the stable transition is
{P-F}. We note also that for $p\sim 0.81$ there can be more
separated such ranges. Finally for $p< 0.81\pm 0.01$ the transition
can only be {P-SG}. Let us now consider the case $J_\mathrm{2}<0$. 
As is evident from
Fig. \ref{f3}, whereas for $J_\mathrm{2}/J_\mathrm{1}>-1$
for any value of $p$ the system may have both a {P-F} or a {P-SG} transition,
for $J_\mathrm{2}/J_\mathrm{1}<-1$
the system has only a {P-SG} transition
(similarly, for $J_1<0$ the {P-AF} transition may occur only if
$J_\mathrm{2}/J_\mathrm{1}>1$).
Recall that, under the restriction of Eq. (\ref{pp}),
the averages of the couplings have the same sign 
of the parameters $J_1$ and $J_2$.
For $p>0.5$, the two completely different scenarios between the two cases
$J_\mathrm{2}>0$ and $J_\mathrm{2}<0$ reflect the
difference of the average of the frustration over the triangles,
positive in the former and negative in the latter case.
By looking back at the transformations (\ref{PF}), we see
that, concerning the disordered {P-F} transitions, the mapping
consists in a simple renormalization of the couplings $J_1$ and $J_2$
so that the same argument used in Sec. 3 to describe the
frustrated regions and, in particular, the dimers-triangles transition,
applies again. On the other hand, unlike the non random case,
for $J_2/J_1<-1$ there exists always a stable and unique {P-SG} transition,
signaling the onset of overlaps orderings.
It is interesting to observe that a picture in terms of non interacting
dimers can be found even in the glassy phase 
in the limit $|J_2|\gg |J_1|$ with $J_1$ kept fixed.
In such a limit in fact, the spins over
a dimer must be exactly parallel or antiparallel, for $J_{x,y}^{(2)}>0$ or 
$J_{x,y}^{(2)}<0$, respectively, regardless of the values
of the other couplings $J_{x,y}^{(1)}$, so that the average over the disorder
of the local magnetization 
of any spin gives zero.

\begin{figure}
\begin{center}
\includegraphics[width=0.6\columnwidth,clip]{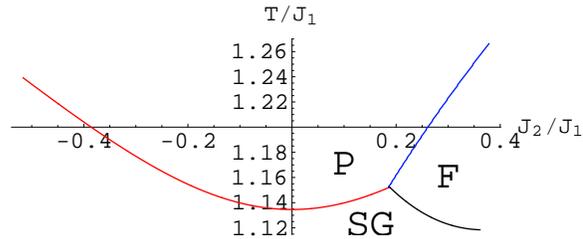}
\end{center}
\caption{ Phase diagram for the case $p_\mathrm{1}=p_\mathrm{2}=0.83$
in the neighborhood of the multicritical point P-F-SG. The line drawn between
the SG and the F phase is the Nishimori line which, as is known, passes
inside the F phase. 
}
\label{f4}
\vspace{0.7cm}
\end{figure}

In Fig. \ref{f4} we report the detail of the phase diagram
of the case $p=0.83$ providing even the neighborhood of the {F-SG} crossover.

Finally, we point out that, as it must be, for any choice of $p$,
the multicritical point F-P-SG belongs to the Nishimori line.
This fact can be immediately checked by substituting 
$\tanh(\beta J)=2p-1$ into Eqs. (\ref{PF}) and (\ref{PSG}). 

\subsubsection{The general case.}
\begin{figure}
\begin{center}
\includegraphics[width=0.6\columnwidth,clip]{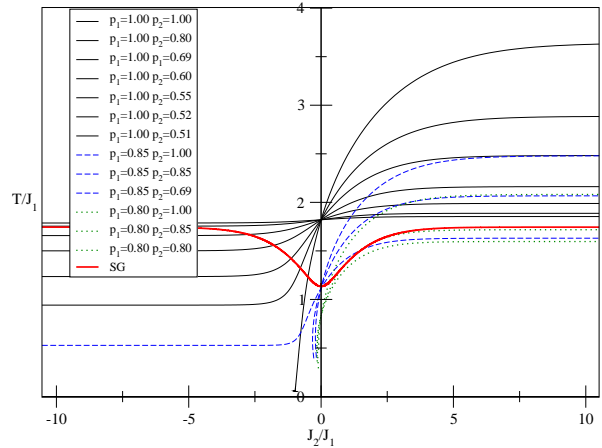}
\end{center}
\caption{ Critical lines of the disordered transition {P-F} for several
values of $p_\mathrm{1}$ and $p_\mathrm{2}$ and the {P-SG} line.
}
\label{f5}
\end{figure}

As shown in Fig. \ref{f5}, for arbitrary values of the parameters
$p_\mathrm{1}$ and $p_\mathrm{2}$ the scenario of the possible
transition lines becomes of course reacher. In the region
$J_\mathrm{2}\geq 0$ we do not observe relevant qualitative
differences with respect to the case $p_\mathrm{1}=p_\mathrm{2}$.
We see that in general, for any fixed value of $p_\mathrm{1}$, by
varying $p_\mathrm{2}$ one obtains a family of non intersecting
critical lines margins at $J_\mathrm{2}=0$ (as given by Eq.
(\ref{PSG0})) and whose $T_c$ in general becomes lower as
$p_\mathrm{2}$ decreases. For fixed values of $p_\mathrm{1}$ and
$p_\mathrm{2}$, as the value of $J_\mathrm{2}$ increases, the
{P-F} lines reach a horizontal asymptote whose value can be
explicitly calculate from Eq. (\ref{PFG})
\begin{eqnarray}
\label{Jginfty}
\frac{T_c^{(\mathrm{F})}(J_\mathrm{2}\to +\infty)}{J_\mathrm{1}}=
\left[\Arctanh\left(\frac{\kappa(B)}{A}\right) \right]^{-1},
\end{eqnarray}
where $A=2p_\mathrm{1}-1$ and $B=2p_\mathrm{2}-1$.
From Fig. \ref{f5} we observe also that two critical lines having two different values of
$p_\mathrm{1}$, in general, may intersect each other.
In the region $J_\mathrm{2}<0$ we observe that if $p_\mathrm{1}$ is sufficiently high,
(starting from $p_\mathrm{1}\sim 0.85$) and $p_\mathrm{2}\neq p_\mathrm{1}$,
as $J_\mathrm{2}$ decreases, the transition lines can have an inversion
and reach another horizontal asymptote whose value is given by
\begin{eqnarray}
\label{Jgneg}
\frac{T_c^{(\mathrm{F})}(J_\mathrm{2}\to -\infty)}{J_\mathrm{1}}=
\left[\Arctanh\left(\frac{\kappa(-B)}{A}\right) \right]^{-1}.
\end{eqnarray}
Note that the temperatures given by this equation are lower than
the ones obtained in the opposite limit $J_\mathrm{2}\to +\infty$.
Note also that Eqs. (\ref{Jginfty}) and (\ref{Jgneg}) have no
solution for values of $p_\mathrm{1}$ and $p_\mathrm{2}$ such that
$A^2B-2AB-2A+1=0$ and $A^2B-2AB+2A+1=0$, respectively. But
analogously to the case previously seen
($p_\mathrm{1}=p_\mathrm{2}$), as evident from Fig. \ref{f5},
these regions of non existence occur only for values of
$p_\mathrm{1}$ and $p_\mathrm{2}$ quite lower than the threshold
at which the only stable transition is {P-SG}.

\section{Conclusions}
The phase diagram of a competitive Ising model
defined over the triangular Husimi tree has been analyzed,
both in the non random and the random version (spin glass model).
First, we have analyzed the non random case providing, in particular,
an interesting physical characterization of the ground states 
in the frustrated regions in terms of a zero temperature phase transition between
an ordered or antiordered state of connected triangles ({F} or {AF})
and a gas of non interacting two state dimers (also called a spin liquid in
the literature). A natural question is whether or not this zero temperature
phase transition represents the zero temperature boundary of a finite
temperature phase transition. 
Second, we have introduced the disorder and studied
the upper phase diagram and the corresponding critical behavior 
of the random case which, in particular, gives an affirmative 
answer to the above question. 
Our method relies on a recently proposed
mapping valid for a wide family of graphs turning out to be
infinite dimensional in a broad sense, as happens (but not only)
in generalized tree-like graphs and in suitable structures where
an infinite number of closed paths per vertex is also allowed.
This second class includes, in particular, generalized Bethe lattices 
and Husimi trees, \textit{i.e.}, graphs over which 
non random models are exactly solvable.
The mapping maps the spin glass model onto the corresponding non random one.
For the two couplings $J_\mathrm{1}$ and
$J_\mathrm{2}$ we have chosen
the disorder giving rise to  
the triangular $\pm J$ model.
We can observe directly the role of the explicit
frustration induced by the presence of the loops tuned by the
amplitude and the sign of the parameter-coupling $J_\mathrm{2}$.
As expected from an argument based on high temperature expansion and on
frustration, the spin glass critical temperature
is a nondecreasing function of
$|J_\mathrm{2}|$, whereas the disordered ferromagnetic critical temperatures
is a nondecreasing function of $J_\mathrm{2}$.
We observe the following scenario.
In the non frustrated regions (that is the sectors I and III of the phase
diagrams), in the spin glass model we find an ``ordinary'' 
competition (like in the Sherrington Kirkpatrick model) between a
{P-SG} and a {P-F} or {P-AF} transitions, in which the parameters $J_1$ and
$J_2$ are simply renormalized by the mapping and the spin glass transition
signals the onset of spin overlaps.
In the frustrated regions instead (sectors II and IV of the phase diagrams), 
we find a new feature; in fact, even for $|J_2|>|J_1|$,
there exists always a phase transition which 
turns out to be a {P-SG} phase transition 
(the red line in the Figs. 8-11), whose glassy phase
consists again of a gas of non interacting two state dimers.

The analytical study carried out here
concerns a non trivial model defined
over a structure where loops are important.
Other more complex and variegated cases, such as other Ising models with
more than two couplings and Potts models, could be similarly analyzed.

Even if, as stressed in Sec. I, in the Husimi tree, 
we do not have the existence of $\varphi_I$,
the density free energy in the thermodynamic limit,  
which is a necessary condition for the mapping, we think nevertheless that 
the application of the mapping also in this case 
provides important information of the random model.
We stress in fact that in the mapping no ad hoc approximation is introduced. 

\appendix
\section{}
A comment is in order concerning the model
considered in the Ref. \cite{GPW1} (and partially in \cite{FarI,FarII}),
which is defined on the finite Husimi tree of Fig. \ref{f0},
and the model given in Sec. 2.1, which is defined on the
corresponding infinite Husimi tree (which can be seen as a
Bethe lattice with loops).

Whereas in the infinite Husimi tree one has to solve the
model with the Hamiltonian (\ref{compet}) for vertices far from
the boundary, and this is done by considering in all the
calculation the lattice as infinite from the very beginning, in
the finite Husimi tree one has to take into account also the boundary
in the finite lattice and to perform the thermodynamic limit in
the proper manner by calculating the correlation functions only from the
density free energy after the thermodynamic limit has been taken.
The relations between infinite and finite Husimi trees are on the
same level of those between Bethe lattices and Cayley trees.
It turns out that the physics, and in particular the
free energy, described by these two versions of models is quite different
\cite{Eggarter,Muller,Peruggi}.
Nevertheless, the equations defining the effective fields for the central
spin used in both
the approaches are the same implying that also the fixed points,
and then the equations for the
critical surfaces and the magnetization of the central spin, are the same.
We show this now by deriving the equations for the infinite Husimi tree
following the so called Baxter approach \cite{Baxter}, subsequently
adapted to generalized Bethe lattices by many authors (see \cite{SC}
and references therein) and to Husimi trees by \cite{Monroe}.

Let us indicate with $Z_n(\sigma_0)$ the conditional partition function
of the root spin $\sigma_0$ (see Fig. \ref{f0}) 
for the $n$-th generation graph;
that is the graph of Fig. \ref{f0} having $n$ shells completed, having
$|S_n|$ vertices, plus an external circular annulus $C_n$, having the latter
a number $|C_n|$ of vertices such that $|S_n|/|C_n|\to 0$ as $n\to\infty$.
We have
\begin{eqnarray}
\label{Zn}
Z_n=\sum_{\sigma_0}Z_n(\sigma_0),
\end{eqnarray}
where $Z_n$ is the full partition function of the $n$-th generation graph.
By looking at Fig. \ref{f0}, it is easy to see that the conditional partition
function $Z_{n+1}(\sigma_0)$ of the root spin at the $n+1$-th generation graph
is related to $Z_n(\sigma_0)$ through \small
\begin{eqnarray}
\label{Znn}
Z_{n+1}(\sigma_0)&=&\sum_{\sigma_1,\sigma_2}
e^{-B\sigma_0+k_1\sigma_0(\sigma_1+\sigma_2)+k_2\sigma_1\sigma_2} Z_n(\sigma_1)Z_n(\sigma_2),
\end{eqnarray}
\normalsize
where $B=\beta h$, $k_1=\beta J_1$, $k_2=\beta J_2$, and $\sigma_1$ and
$\sigma_2$ are the two Ising variables located at the vertices 1 and 2
respectively in Fig. \ref{f0}.
It is convenient to introduce the effective field $h_n$: 
\begin{eqnarray}
\label{hn}
e^{2h_n}=\frac{Z_n(-)}{Z_n(+)}.
\end{eqnarray}
Performing the sums over $\sigma_1$ and
$\sigma_2$ in Eq. (\ref{Znn}) both for $\sigma_0=+$ and $\sigma_0=-$, and using
the definition (\ref{hn}) we arrive at the following closed recursive
equation for the effective field \small
\begin{eqnarray}
\label{hnn}
e^{2h_{n+1}}=e^{2B}\frac{e^{-2k_1+k_2}+2e^{-2k_2+2h_n}+e^{2k_1+k_2+4h_n}}
{e^{2k_1+k_2}+2e^{-2k_2+2h_n}+e^{-2k_1+k_2+4h_n}}.
\end{eqnarray}
\normalsize
All the possible
phase transitions (except the modulated ones) 
of the uniform model can be studied analyzing the
fixed points $h^{*}$ of the above equation: \small
\begin{eqnarray}
\label{h*}
e^{2h^{*}}=e^{2B}\frac{e^{-2k_1+k_2}+2e^{-2k_2+2h^{*}}+e^{2k_1+k_2+4h^{*}}}
{e^{2k_1+k_2}+2e^{-2k_2+2h^{*}}+e^{-2k_1+k_2+4h^{*}}}.
\end{eqnarray}
\normalsize
Of course one can use Eq. (\ref{h*}) to write the recursive equation for
the magnetization for the root spin $\sigma_0$.
If one is interested in studying the individual magnetizations of further spins
as for the spins $\sigma_1$ and $\sigma_2$, then one has to write the
recursive equations for the conditional partition function
$Z_{n+1}(\sigma_0,\sigma_1,\sigma_2)$ in terms of weighted sums over
$\sigma_{1,1},\ldots \sigma_{2,2}$ of the conditional partition functions
$Z_{n}(\sigma_1,\sigma_{1,1},\sigma_{1,2})$ and
$Z_{n}(\sigma_2,\sigma_{2,1},\sigma_{2,2})$ (see Fig. \ref{f0}) and defining
the corresponding effective fields. In this way one can study possible
modulated phases as done for example in \cite{SC} for generalized Bethe
lattices involving second and third-nearest neighbor interactions.

In the model considered in this paper, the base spin $\sigma_0$ is attached
to one single triangle (see Fig. \ref{f0}). However one could be interested in
studying cases in which the root spin has to be attached to more, say $q$,
triangles. It is immediate to recognize that in this case instead of
Eq. (\ref{h*}) one has 
\begin{eqnarray*}
\label{h*q}
e^{2h^{*}}=e^{2B}\left[
\frac{e^{-2k_1+k_2}+2e^{-2k_2+2h^{*}}+e^{2k_1+k_2+4h^{*}}}
{e^{2k_1+k_2}+2e^{-2k_2+2h^{*}}+e^{-2k_1+k_2+4h^{*}}}
\right]^{q}.
\end{eqnarray*}
The model considered in this paper corresponds to the case $q=1$ of
Eq. (\ref{h*}) which was also derived in \cite{GPW1} for the finite
Husimi tree. As already mentioned, despite the physics for the infinite and
finite Husimi trees be different, the equations for the effective fields are
the same, or in other words, the physics of the ``central'' spins, far from
the boundary, are the same.
Therefore we can, in particular, use
the analysis of the Ref. \cite{GPW1} for the phase transitions
consisting in the study of the solutions of Eq. (\ref{h*}) and their
stability leading to the Eqs. (\ref{F}) and (\ref{AF}).


\section*{Acknowledgments}
This work was supported by the FCT (Portugal) grants
SFRH/BPD/17419/2004, SFRH/BPD/24214/2005, pocTI/FAT/46241/2002 and
pocTI/FAT/46176/2003, and the Dysonet Project.
We thank A. V. Goltsev and M. Hase, 
for many useful discussions and a critical reading
of the manuscript.


\begin{thebibliography}{10}
\bibitem{Nota}
Due to this similarity with a glassy phase,
we could alternatively use the expression ``liquid of dimers'' 
instead of ``gas of dimers'' for this phase. 
Of course such a distinction for our model is just a lexical fact.
\bibitem{Nota1}
In the Ref. \cite{MOI} was claimed that
the probability for the overlap $p(l)$ has to go zero for $l\to\infty$.
Of course such a relation has to be satisfied if $p(l)$ is a probability.
The above statement remains sensible but only heuristic.
In fact, more precisely, it is necessary to define the probability
for two finite paths of lenght $l_1$ and $l_2$ to overlap $l\leq l_1,l_2$ 
times, $p_{l_1,l_2}(l)$, and to analyze its behavior for $l_1,l_2\to\infty$.
It is possible then to show rigorously that the uniform bound 
$p_{l_1,l_2}(l)<C\exp(-bl)$, with $C$ and $b$ two positive constant,
is a sufficient condition for the mapping to become exact \cite{MOIII}.
Note that, in general, such a bound does not guarantee the existence
of an asymptotic probability $p(l)$. 
However, in the case of Bethe lattices or Husimi trees (and in many random
graphs) one has also the existence
of $p(l)=\lim_{l_1,l_2\to\infty}p_{l_1,l_2}(l)$, so that the sufficient
condition can be directly verified looking at the distribution $p(l)$.
\end{thebibliography}
\end{document}